# Fourier Transform Model for All-Order PMD Compensation based on a Coupled-Mode Equation Solution using the First Born Approximation

Michael C. Parker (1), Etienne Rochat (2), and Stuart D. Walker (2)

1: Fujitsu Network Communications Inc., Photonics Networking Laboratory, Colchester, CO3 4HG, UK,
   Tel: +44(0)1206 542399, Fax: +44(0)1206 762916, e-mail: M.Parker@ftel.co.uk
2: University of Essex, Department of Electronic Systems Engineering, Wivenhoe Park, Colchester CO4 3SQ, UK,

**Abstract** *We present a Fourier transform methodology for all-order PMD analysis, based on the first Born approximation to the coupled-mode equation solution. Our method predicts wavelength-dependent PMD effects and allows design of filters for their mitigation.*

## Introduction

Polarization mode dispersion (PMD) has been extensively studied [1-6], and can be characterized to first order in terms of differential group-delay (DGD-1) in the fiber principle states of polarization (PSP-1). Second order effects such as polarization-state rotation or depolarization (DEP-2), and polarization chromatic dispersion (PCD-2) are of concern, as 40Gb/s channel data rates become commercial reality. In this paper, we show how high-order PMD trajectories on the Poincaré sphere surface [7] can be understood using coupled-mode theory. A Fourier transform (FT) basis allows a scaleable filter method, shown below in figure 1, which enables compensation of all PMD orders on both a static and dynamic basis.

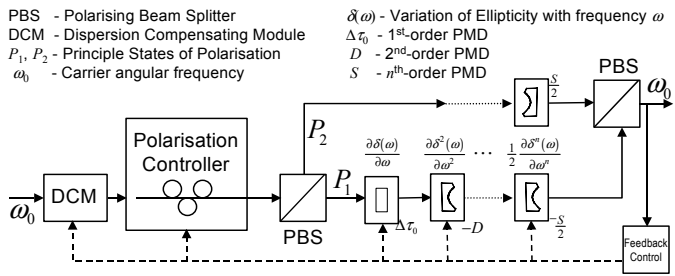

PBS - Polarising Beam Splitter  
DCM - Dispersion Compensating Module  
$P_1, P_2$ - Principle States of Polarisation  
$\omega_0$ - Carrier angular frequency  
$\delta(\omega)$ - Variation of Ellipticity with frequency $\omega$  
$\Delta\tau_0$ - 1st-order PMD  
$D$ - 2nd-order PMD  
$S$ - $n$th-order PMD  

Fig.1: Schematic of all-order PMD compensation, via consideration of variation in polarization ellipticity $\delta$ with angular frequency $\omega$.

## Theory

Generally, PMD is not a problem of signal energy loss, but of dispersion. A signal is characterised by its amplitude and phase, as they vary in the temporal and frequency domains. Classically, the energy of a signal is only a function of the amplitude, so if a signal suffers distortion but no energy loss, that distortion only arises due to phase variation. Hence, PMD must be a phase problem, which is correctable using phase-only compensation techniques, such as all-pass filters (APFs) [8]. We note that for a causal signal, the amplitude and phase must be related via a Hilbert transform, so that phase variation must imply an amplitude variation, and hence energy loss. But, due to its relative unimportance, we neglect this aspect in the following analysis. PMD is caused by energy coupling between the two non-degenerate polarisation modes of standard SMF, and hence it can be analysed using coupled-mode theory. Using the notation of Yariv [9], we write:

$$\partial_z P_1 = -j\kappa(z) P_2 e^{j\Delta\beta z} \quad (1), \quad \partial_z P_2 = -j\kappa^*(z) P_1 e^{-j\Delta\beta z} \quad (2)$$

where $P_1$ and $P_2$ are the electric field amplitudes in the two local principle states of polarisation (PSPs), $z$ is the longitudinal coordinate, $\Delta\beta = \beta_1 - \beta_2$ is the difference between the propagation constants associated with each of the PSPs, where $\beta_{1,2} = 2\pi n_{1,2}/\lambda$ and $n_{1,2}$ are the local refractive indices associated with each of the PSPs. The strength of localised coupling between the two PSPs is closely of the form:

$$\kappa(z) = \frac{-j}{2\Delta n(z)} \frac{\partial \Delta n(z)}{\partial z} e^{j\phi(z)} \quad (3)$$

where $\Delta n(z)$ is the local birefringence in the fibre, causing the mode coupling, and $\phi(z)$ is the appropriate phase change. We note equation (3) indicates that both high and low birefringences have small coupling coefficients, which is to be expected for HiBi (polarisation maintaining) fibre, and perfectly degenerate SMF respectively. This leaves the general case of standard non-degenerate SMF, where polarisation mode-coupling is an important issue. Since the PMD coupling is considered weak, we can employ the first Born approximation to the solution of the coupled-mode equations to yield the scattering probability for coupling between modes 1 and 2:

$$\rho_{1,2} = -j \int_{-\infty}^{\infty} \kappa^*(z) e^{-j\Delta\beta z} dz \quad (4)$$

The evolution of coupling coefficient $\kappa(z)$ along the length of the SMF follows a random walk. This stochastic property is preserved by the Fourier integral of (4), to yield an equivalent random walk in the spectral domain. Scattering amplitude between modes 2 and 1 is $\rho_{2,1} = -\rho_{1,2}^*$, whilst the probability of no scattering (i.e. polarisation maintenance) is given from $|\tau| = \sqrt{1 - |\rho|^2}$. With no losses or polarisation-dependent loss (PDL) assumed, we can write the unitary frequency-dependent scattering matrix as:

$$\begin{bmatrix} P_1 \\ P_2 \end{bmatrix}_{out} = \begin{bmatrix} \tau(\omega) & \rho(\omega) \\ -\rho^*(\omega) & \tau^*(\omega) \end{bmatrix} \begin{bmatrix} P_1 \\ P_2 \end{bmatrix}_{in}. \quad (5)$$

Given that the input signal of carrier frequency $\omega_0$ is aligned to the SMF PSP at that frequency, then we have that $\begin{bmatrix} P_1(\omega_0) & P_2(\omega_0) \end{bmatrix}_{in}^T = \begin{bmatrix} 1 & 0 \end{bmatrix}^T$, and hence $\begin{bmatrix} P_1(\omega) & P_2(\omega) \end{bmatrix}_{out}^T = \begin{bmatrix} \tau(\omega) & -\rho^*(\omega) \end{bmatrix}^T$. Assuming the same intensity, an elliptical state of polarization (SOP) is a unique point on the surface of the Poincaré sphere [10], described by two spherical angular coordinates: $\delta$ the ellipticity, and $\chi$ the orientation of the ellipse. Higher-order PMD is characterized by trajectories on the surface of the Poincaré sphere, which describe the variation with angular frequency of ellipticity and orientation of the SOP, $\delta(\omega)$ and $\chi(\omega)$ respectively. Using a polarization controller, it is possible to align the output PSP of the SMF, defined at the carrier frequency $\omega_0$, to that of a polarizing beam splitter (PBS), whose orientation is $\chi = 0$. When an arbitrary polarization is passed through a PBS, if the orientation is aligned to that of the PBS (i.e. $\chi(\omega_0) = 0$), then the elliptical polarization is simply resolved into its component major and minor axis amplitudes, $a_1$ and $a_2$. However, the ellipticity angle $\delta$ is conserved between the two linearly-polarized components, and manifests itself as an appropriate complex exponential phase. The electric fields in the orthogonal PSPs, $P_1$ and $P_2$, are directed respectively into the two arms of the Mach-Zehnder interferometer (MZI) formed by two PBSs, as shown in figure 1. When the trajectory of the depolarization is such that $\chi$ varies with frequency and is no longer necessarily zero, then the amplitudes of the Jones matrix elements are modified and given by:

$$P_1(\omega)_{out} = \begin{bmatrix} a_1(\omega) \\ a_2(\omega) \end{bmatrix} e^{j\delta(\omega)/2}, \qquad (6a)$$

$$P_2(\omega)_{out} = \begin{bmatrix} -a_1(\omega) \\ a_2(\omega) \end{bmatrix} e^{-j\delta(\omega)/2}. \qquad (6b)$$

Aligning the PBS orientation to $\chi = 0$, we must have $a_1(\omega_0) e^{j\delta(\omega_0)/2} \equiv \tau(\omega_0)$, and $a_2(\omega_0) e^{-j\delta(\omega_0)/2} \equiv -\rho^*(\omega_0)$. Thus the PBS acts to convert an arbitrary polarization into two linear polarizations each with a conjugate phase change. However, from filter theory, it is the phase response (i.e. variation of phase with frequency), which determines the dispersion characteristic. A linear phase response corresponds to a uniform group delay for all frequencies, but any departure from linearity manifests itself as multi-order chromatic dispersion. Thus the group-delay $\Delta \tau_0$ between the two PSPs, and their respective 2nd-order chromatic dispersions $D$ are given by [11]:

$$\Delta \tau_0 = -\frac{\partial \delta(\omega)}{\partial \omega} \quad (7a) \qquad D = -\frac{\partial^2 \delta(\omega)}{\partial \omega^2}. \quad (7b)$$

Group-delays of $+\Delta\tau_0/2$ and $-\Delta\tau_0/2$ appear in the respective arms of the MZI, and can be considered to be the 1st-order PMD, DGD-1; and simply compensated by a single delay line of magnitude $\Delta\tau_0$ in one of the arms. Conjugated second- and higher-order chromatic dispersions appear in each of the arms of $\pm D/2$ and $\pm S/2$, respectively. These can be considered to be the 2nd-order (PCD-2 and DEP-2) and higher-order PMD terms. For each PSP, the two arms of the MZI separately compensate for the delay and higher-order dispersions, and the 2nd PBS brings them together, to yield a superposed compensated signal. $\chi(\omega)$ simply manifests itself as a delay and chromatic dispersion (with higher-orders) common to both arms of the MZI, essentially independent of polarization. The delay can be ignored, whilst the chromatic dispersion associated with $\partial^2\chi(\omega)/\partial\omega^2$ can be compensated via a feedback loop to the dispersion compensation module (DCM) immediately before the polarization controller. The DCM is essentially there to compensate the material dispersion of the SMF link. However, by increasing the dispersion of the DCM by a further $+D/2$, 2nd-order PMD can be compensated by a single active dispersion compensator of magnitude $-D$ in only one of the arms, as indicated in figure 1. Higher-order dispersions [12] (without a common compensator before the MZI) require phase-conjugated dispersion compensators for full compensation, as indicated. Optical lattice filters would be ideal candidates for this type of application.

**Conclusions**
We have described a complete FT model for PMD effects using the first Born approximation technique for solution of the coupled-mode equations. Not only does it allow PMD effects to be modelled by standard signal theory, but appropriate mitigating filter designs can also be synthesised. In addition, we have shown how polarization trajectories on the Poincaré sphere surface due to PMD effects can be understood from coupled-mode theory.